\documentclass[prl,amsmath,aps,twocolumn]{revtex4}
\usepackage{amssymb}
\usepackage{times}
\usepackage{graphicx}
\newcommand{\tr}{{\rm Tr}}

\newcommand{\mX}{{\mathcal X}}
\newcommand{\mZ}{{\mathcal Z}}
\newcommand{\mS}{{\mathcal S}}
\newcommand{\mU}{{\mathcal U}}
\newcommand{\mV}{{\mathcal V}}
\newcommand{\mG}{{\mathcal G}}
\newcommand{\mM}{{\mathcal M}}
\newcommand{\mI}{{\mathcal I}}
\newcommand{\mA}{{\mathcal A}}
\newcommand{\mB}{{\mathcal B}}
\newcommand{\mT}{{\mathcal T}}
\newcommand{\mP}{{\mathcal P}}
\newcommand{\mO}{{\mathcal O}}
\newcommand{\mh}{{\mathbf h}}
\newcommand{\md}{{\mathbb D}}

\begin{document}

\title{Two infinite families of nonadditive quantum error-correcting codes}
\author{Sixia Yu$^{1,2}$, Qing Chen$^{1,2}$, and C. H. Oh$^2$}
\affiliation{$^1$Hefei National Laboratory for Physical Sciences
at Microscale and Department of Modern Physics \& Department of
Modern Physics University
of Science and Technology of China, Hefei 230026, P.R. China\\
$^2$Physics Department, National University of Singapore, 2 Science
Drive 3,  Singapore 117542}
\begin{abstract}
We construct explicitly two infinite families of genuine nonadditive
1-error correcting quantum codes and prove that their coding
subspaces are 50\% larger than those of the optimal stabilizer codes
of the same parameters via the linear programming bound. All these
nonadditive codes can be characterized by a stabilizer-like
structure and thus their encoding circuits can be designed in a
straightforward manner.
\end{abstract}
\maketitle

One major family of quantum error-correcting codes (QECCs)
\cite{shor1,ben,ste1,knill}, which are powerful tools to fight the
quantum noises in various quantum informational processes, are
called as {\em additive} or {\em stabilizer} codes
\cite{g1,g2,cal1,cal2}. The coding subspace of a stabilizer code is
specified by the joint +1 eigenspace a group of commuting multilocal
(direct product of) Pauli operators. Usually $[[n,k,d]]$ denotes a
stabilizer code of length $n$, the number of physical qubits, and
distance $d$, i.e., correcting up to
$\lfloor\frac{d-1}2\rfloor$-qubit errors, that encodes $k$ logical
qubits ($2^k$-dimensional subspace).

The first example of {\em nonadditive} codes, codes that cannot be
described within the framework of stabilizer, was an infinite family
of 1-error-detecting codes \cite{rains,rainsa}, e.g., $((5,6,2))$,
with coding subspaces being 50\% larger than the optimal stabilizer
codes of the same parameters. Recently another family of 1-error
detecting codes with still larger encoding subspaces has been
constructed in \cite{smolin} and slightly improved in \cite{yu2}.
Here we have denoted by $((n,K,d))$ a nonadditive code of length $n$
and distance $d$ that encodes a $K$-dimensional logical subspace
(about $\log_2K$ logical qubits).

The first example of nonadditive code \cite{yu1}, namely
$((9,12,3))$, that outperforms all  the stabilizer codes of the same
length while capable of correcting arbitrary single qubit errors has
recently been constructed via a graphical approach based on graph
states. Later on an optimal 10-qubit code $((10,24,3))$ has been
found via a comprehensive computer search \cite{yu2}. Recently a
family of codes of distance 8 that encode 3 more logical qubits than
the best known stabilizer codes have been constructed based on
 nonlinear classical codes \cite{grsl}. However the possibility of
being equivalent to some subcode of an optimal stabilizer code or
even a stabilizer code of the same parameters under local unitary
transformations has not yet been excluded.

Generally, being without a stabilizer structure, the nonadditive
codes promise larger coding subspaces while they are harder to
construct and identify than the stabilizer codes. On one hand  there
is no systematic construction so far and all the good codes are
found via computer search \cite{yu2,zb}, which is impossible for a
relatively large length (e.g. $n\ge 11)$.  On the other hand an
obvious criterion for a genuine nonadditive code is to check whether
or not its coding subspace is larger than all the stabilizer codes
of the same parameters. However the exact bound for stabilizer codes
is generally unknown. As a result it is of interest to find
nonadditive error-correcting code when the length tends to infinity
that outperforms all the stabilizer codes of the same parameters.

In this Letter we shall construct two infinite families of {\it
genuine} nonadditive 1-error-correcting codes with coding subspaces
being 50\% larger then the corresponding {\it optimal}
1-error-correcting stabilizer codes of the same parameters to show
that the nonadditive error-correcting codes outperform the
stabilizer codes even when the length $n$ tends to infinity. All the
nonadditive codes are characterized by a stabilizer-like structure
and therefore the encoding-decoding circuits can be designed in a
straightforward manner.

Two families of nonadditive 1-error correcting codes that we shall
construct have the following parameters
\begin{equation}\label{d}
\mathbb D_{(m,a)}=((N^a_m,\textstyle\frac 32 2^{N_m^a-2m-6},3))
\end{equation}
where $N_m^a=\frac{2^{2m+5}-5}3+a$ with $a=0,1$ and $m\ge 1$. To
ensure that they are genuine nonadditive we shall prove that the
corresponding optimal 1-error-correcting stabilizer codes of the
same length have parameters
\begin{equation}\label{os}
[[N_m^a,N_m^a-2m-6,3]]
\end{equation}
by working out analytically the linear programming bound for the
lengths $N_m^a$. Notice that the quantum Hamming bound permits
exactly one more logical qubit, i.e., $\lceil
\log_2(3N_m^a+1)\rceil=2m+5$. The first nonadditive codes of these
two families are $((41,3\cdot 2^{32},3))$ and $((42,3\cdot
2^{33},3))$ respectively while the optimal stabilizer codes have the
parameters $[[41,33,3]]$ and $[[42,34,3]]$.

Our construction is based on a family of stabilizer codes of lengths
$\{2^{2r+3}\}_{r=1}^{m}$ \cite{g2, cal2} and two nonadditive codes
of length 9 and 10 discovered recently \cite{yu1,yu2}  and is a kind
of pasting stabilizer codes with nonadditive codes that generalizes
the pasting of stabilizer codes in Ref.\cite{g3}. We denote by
$\mX_v,\mZ_v$ three Pauli operators acting nontrivially only on some
qubit labeled by $v$ and by $\mI$ the identity operator. Furthermore
for a given index set $U$ we denote $\mX_U=\prod_{v\in U}\mX$ and
similarly for other Pauli operators.

Let us look at the optimal stabilizer of length $2^{2r+3}$ at first.
According to Ref.\cite{g2} the stabilizer of the code has $2r+5$
generators with two of them being $\mX_{U_r}$ and $\mZ_{U_r}$ where
we have labeled $2^{2r+3}$ physical qubits with
$U_r=\{1,2,3,\ldots,2^{2r+3}\}$. The remaining $2r+3$ generators are
given by
\begin{equation}\label{s}
\left\{\mS^r_k=\mX^{\mh_k}\mZ^{\mh_{k-1}+\mh_1+\mh_{2r+3}}\mid k\in
U_r\right\}.
\end{equation}
Here $\mh_k$ denotes a $2^{2r+3}$-dim vector that is the $k$-th row of a $(2r+3)\times 2^{2r+3}$ matrix $H_r=[c_0,c_1,\ldots,c_{2^{2r+3}-1}]$ whose $k$-th
column $c_k$ being the binary representation of $k$, e.g., $c_1^T=(0,0,\ldots,1)$
and $c_{2^{2r+3}-1}=(1,1,\ldots,1)$ and $\mh_0=\mathbf 0$ is the zero vector. And for a vector $\mh$ with components $\{h_v\mid v\in U_r\}$ we have denoted
$\mX^\mh=\prod_{v\in U_r}\mX_v^{h_v}$ and $\mZ^\mh=\prod_{v\in U_r}\mZ_v^{h_v}$.

Despite their nonadditiveness the codes $((9,12,3))$ and
$(10,24,3))$ admit a stabilizer-like structure and can be most
conveniently formulated by using the graph states
\cite{graph,werner}. We denote by $G=(V,E)$ a simple undirected
graph with a set $V$ of vertices and a set $E$ of edges. Two
vertices are connected with an edge iff $\{a,b\}\in E$. Two graphs
$G_a$ $(a=0,1)$ on $|V_0|=9$ and $|V_1|=10$ vertices are shown in
Fig.1. By labeling $|V|$ qubits by $V$ we can define the graph state
corresponding to the graph $G$ a $|G\rangle=\mU_G|+\rangle_x^V$
where
\begin{equation}
\mU_G=\prod_{\{a,b\}\in E}\frac{1+\mZ_a+\mZ_b-\mZ_a\mZ_b}2,
\end{equation}
and $|+\rangle^V_x$ denotes the joint +1 eigenstate of $\mX_v$ for
$v\in V$. Obviously $\mU_G^2=1$ and the graph state $|G\rangle$ is
also the +1 joint eigenstate of the following $n$ stabilizers
$\mG_v=\mU_G\mX_v\mU_G$.

\begin{figure}[b]
\includegraphics{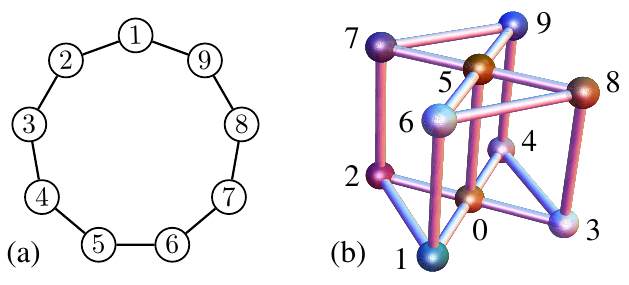}
\caption{Two graphs $G_a=(V_a,E_a)$ are shown with
$V_0=\{1,2,\ldots,9\}$ and $V_1=\{0,1,\ldots,9\}$. They provide the
graph-state bases for two nonadditive codes $((9,12,3))$ and
$((10,24,3))$.}
\end{figure}

From the graph state $|G\rangle$  a basis for the whole system a
basis of the whole system can built $\{\mZ_C|G\rangle\mid C\subseteq
V\}$. For trivial graph with no edge the graph-state basis reads
$\mZ_C|+\rangle^V_x$. Any given collection of vertex subsets of $V$
will define a basis that spans a subspace which is referred to as
the graph-state basis. With a graph and a collection of vertex
subsets will define a subspace.

For the nonadditive code $((9,12,3))\equiv\md_{(0,0)}$ we consider
the loop graph $G_0=(V_0,E_0)$ on 9 vertices as shown in Fig.1a and
corresponding graph state $|G_0\rangle$.  The joint +1 eigenspace of
the following 6 observables
\begin{subequations}\label{A}
\begin{eqnarray}
\alpha_1&=&\mU_{G_0}\mX_{\{3,8\}}\mU_{G_0},  \\
\alpha_2&=&\mU_{G_0}\mX_{\{6,2\}}\mU_{G_0},  \\
\alpha_3&=&\mU_{G_0}\mX_{\{9,5\}}\mU_{G_0},  \\
\mA_1&=&\mU_{G_0}(\mX_{\{4,7,3,6,9\}}\mV_{69})\mU_{G_0}, \\
\mA_2&=&\mU_{G_0}(\mX_{\{1,7,3,6\}}\mV_{39})\mU_{G_0}  , \\
\mA_3&=&\mU_{G_0}(\mX_{\{1,4,3,9\}}\mV_{36})\mU_{G_0}  ,
\end{eqnarray}
\end{subequations}
is exactly the 12-dim coding subspace of the code $\md_{(0,0)}$.
Here we have denoted $\mV_{ab}=\frac12(1+\mX_a+\mX_b-\mX_{a}\mX_b)$.
If we denote $\mA_0=\mA_1\mA_2\mA_3$ we have $\tr_{V_0}\mA_0=2^8$.
It should be noted that the code is nondegenerate, which can be
easily seen from its weight distributions.

The nonadditive optimal code $((10,24,3))\equiv\md_{(0,1)}$ has a
graph-state basis corresponding to the graph $G_1=(V_1,E_1)$ on 10
vertices as shown in Fig.1b and the corresponding graph state is
$|G_1\rangle$. Obviously the graph is unchanged under the following
two permutations
\begin{equation}
\pi=({14})({23})(69)(78),\quad \tau=({12})({34})({67})({89}).
\end{equation}
that act naturally on  $V$. Two swapping operators $\mM_{\pi}$ and
$\mM_{\tau}$ are defined via, e.g.,
$\mM_\pi\mZ_C|+\rangle^{V_1}_x=\mZ_{\pi(C)}|+\rangle^{V_1}_x$ for an
arbitrary $C\subseteq V$. For later use we define two
controlled-swap operations with qubits 0 and 5 as sources
respectively as
\begin{subequations}
\begin{eqnarray}
\mT_\pi&=&\frac12\big(1+\mX_0+(1-\mX_0)\mM_\pi\big),\\
\mT_\tau&=&\frac12\big(1+\mX_5+(1-\mX_5)\mM_\tau\big),
\end{eqnarray}
\end{subequations}
which can be readily implemented via Toffli and control-not gates.
According to Ref.\cite{yu2}, those 24 graph-state bases of the code
$\md_{(0,1)}$ are given by
$\{|C^i_{\mu\nu}\rangle=\mZ_{C_{\mu\nu}^i}|G_1\rangle\}$ with
$i=1,2,\ldots,6$ and $\mu,\nu=0,1$ where  24 subsets of $V_1$ are
\begin{equation}
C_{\mu\nu}^i=\pi^\mu\circ\tau^\nu( C_i)\bigtriangleup\nu
B\bigtriangleup\mu \tau^\nu(A)
\end{equation}
with $A\bigtriangleup B=A\cup B-A\cap B$ being the symmetric
difference between two sets $A$ and $B$ and
\begin{eqnarray}
\begin{array}{c}
A=\{0,2,3\},B=\{5,1,2\}, C_1=\emptyset, \cr C_2=\{1,2,3,9\},
 C_3=\{1,2,7,8\}, C_4=\{1,2,6,7,9\},\cr C_5=\{1,3,7,8,9\}, C_6=\{1,3,4,6,7,9\}.
\end{array}
\end{eqnarray}

It turns out that the nonadditive code $\md_{(0,1)}$ admits also a
stabilizer-like structure. With the help of the following encoding
operator
\begin{equation}
\mU_{\rm enc}=\mZ_2\mU_{G_1}\mT_\tau\mT_\pi\mZ_2
\end{equation}
it can be readily checked that  the following 6 mutually commuting
observables
\begin{subequations}\label{B}
\begin{eqnarray}
\beta_1&=&\mU_{\rm enc}\mX_{\{2,3,7\}}\mU_{\rm enc}^\dagger, \\
\beta_2&=&\mU_{\rm enc}\mX_{\{6,7,8\}}\mU_{\rm enc}^\dagger, \\
\beta_3&=&\mU_{\rm enc}\mX_{\{3,4,6,9\}}\mU_{\rm enc}^\dagger,\\
\mB_0&=&\mU_{\rm enc}\left(\mX_{{6}}\mV_{67}\right) \mU_{\rm enc}^\dagger,\\
\mB_1&=&\mU_{\rm enc}\left(\mX_{\{1,2\}}\mV_{37}\right)\mU_{\rm enc}^\dagger,\\
\mB_2&=&\mU_{\rm enc}\left(\mX_{4}\mV_{36}\right) \mU_{\rm
enc}^\dagger
\end{eqnarray}
\end{subequations}
stabilize the coding subspace of $\md_{(0,1)}$, i.e., one one hand
$\beta_k|C_{\mu\nu}^i\rangle=\mB_l|C_{\mu\nu}^i\rangle=|C_{\mu\nu}^i\rangle$
for all possible $k,l,i,\mu,\nu$ and on the other hand the joint +1
eigenspace of these 6 observables, whose projector is given by
\begin{equation}
\mP_1=\prod_{i=1}^3\frac{1+\beta_i}2\prod_{i=0}^2\frac{1+\mB_i}2,
\end{equation}
has exactly dimension 24, i.e.,  $\tr_{V_1}\mP=24$ since
$\tr_{V_1}\mB_0=2^9$. An encoding circuit can therefore be designed
in a similar manner as that of $((9,12,3))$ \cite{yu1}. We note also
that this nonadditive code is non-degenerate.

\begin{table}[t]
\renewcommand{\arraystretch}{1.3}
\caption{The stabilizing observables of the nonadditive codes
$\mathbb D_m^a$ whose physical qubits are labeled with $U_m\cup \ldots \cup U_1\cup V_a$ $(a=0,1)$. The blank
entries stand for suitable  identity operators $\mI_{U_k}$ or $\mI_{V_a}$.}
$$\begin{array}{c|cccccc} \hline\hline
&U_{m}&U_{m-1}&\cdots& U_{2}&U_{1}&V_0\ {\rm or} \ V_{1}\cr\hline
\mO_1^{(0,1)}&\mX_{U_m}&&&&&\cr
\mO_2^{(0,1)}&\mZ_{U_m}&&&&&\cr\hline
\mO_3^{(0,1)}&\mS^{m}_{1}&\mX_{U_{m-1}}&&&&\cr
\mO_4^{(0,1)}&\mS^{m}_{2}&\mZ_{U_{m-1}}&&&&\cr\hline
\mO_5^{(0,1)}&\mS^{m}_{3}&\mS^{m-1}_{1}&&&&\cr
\vdots&\vdots&\vdots&\ddots&&&\cr
\mO_{2m-4}^{(0,1)}&\mS^{m}_{2m-6}&\mS^{m-1}_{2m-8}&\cdots&&&\cr\hline
\mO_{2m-3}^{(0,1)}&\mS^{m}_{2m-5}&\mS^{m-1}_{2m-7}&\cdots&\mX_{U_{2}}&& \cr
\mO_{2m-2}^{(0,1)}&\mS^{m}_{2m-4}&\mS^{m-1}_{2m-6}&\cdots&\mZ_{U_{2}} && \cr\hline
\mO_{2m-1}^{(0,1)}&\mS^{m}_{2m-3}&\mS^{m-1}_{2m-5}&\cdots&\mS^2_{1}&\mX_{U_{1}} \cr
\mO_{2m}  ^{(0,1)}&\mS^{m}_{2m-2}&\mS^{m-1}_{2m-4}&\cdots&\mS^2_{2}&\mZ_{U_{1}} \cr\hline
\mO_{2m+1}^{(0,1)}&\mS^{m}_{2m-1}&\mS^{m-1}_{2m-3}&\cdots&\mS^2_{3}&\mS_{1}^1&\alpha_1\ {\rm or} \ \beta_1\cr
\mO_{2m+2}^{(0,1)}&\mS^{m}_{2m}&\mS^{m-1}_{2m-2}&\cdots&\mS^2_{4}&\mS_{2}^1&\alpha_2\ {\rm or} \ \beta_2\cr
\mO_{2m+3}^{(0,1)}&\mS^{m}_{2m+1}&\mS^{m-1}_{2m-1}&\cdots&\mS^2_5&\mS_{3}^1&\alpha_3\ {\rm or} \ \beta_3\cr
\mO_{2m+4}^{(0,1)}&\mS^{m}_{2m+2}&\mS^{m-1}_{2m}&\cdots&\mS^2_{6}&\mS_{4}^1&\mathcal A_1\ {\rm or} \ \mathcal B_1\cr
\mO_{2m+5}^{(0,1)}&\mS^{m}_{2m+3}&\mS^{m-1}_{2m+1}&\cdots&\mS_{7}^2&\mS_{5}^1&\mathcal A_2\ {\rm or} \ \mathcal B_2\cr\hline
\mO_{2m+6}^{(0,1)}&&&&&&\mathcal A_0\ {\rm or} \
\mathcal B_0\cr\hline\hline
\end{array}$$
\end{table}
Now we are ready to present our construction. We consider $N_m^a$
qubits and label them by disjoint set $U_m\cup U_{m-1}\cup\ldots
\cup U_1\cup V_a$ with $|U_k|=2^{2k+3}$ and $|V_a|=9+a$ with $k\le
m$ and $a=0,1$. We claim that the joint +1 eigenspace of those
$2m+6$ observables $\{\mO^{(a)}_i\}_{i=1}^{2m+6}$ with $a=0$ or $1$
as defined in Table I is the code $\mathbb D_{(m,a)}$  in
Eq.(\ref{d}) with the following projector onto the coding subspace
\begin{equation}
\mP_m^a=\prod_{i=1}^{2m+6}\frac{1+\mO_{i}^{(a)}}2.
\end{equation}
In Table I observables $\mS_i^r$ are defined in Eq.(\ref{s}) and
$\{\alpha_i,\mA_j\}$, $\{\beta_i,\mB_j\}$ are defined via
Eq.(\ref{A}) and Eq.(\ref{B}) respectively. Blank entries represent
suitable identity operators. By juxtaposition of some operators in
the same row we mean their direct product.

First of all, these $2m+6$ stabilizing observables detect all
2-qubit errors because firstly all errors happened on $U$-blocks or
$V$ blocks can be detected because all the subcodes are pure 1-error
correcting codes and secondly all two errors happened on different
qubit blocks can always be detected by the stabilizer containing
$\mX_{U_k}$ and $\mZ_{U_k}$ for some $k$. Thus we obtain a pure
1-error correcting codes of length $N_m^a$. Secondly, by noticing
$\tr \mO_i^{(a)}=0$ for $i\le 2m+5$ and $\tr
\mO^{(a)}_{2m+6}=2^{N_m^a-1}$ we have
\begin{equation}
\tr\mP_m^a=\frac{\tr(1+\mO^{(a)}_{2m+6})}{2^{2m+6}}=\frac32
2^{N^a_m-2m-6}.\end{equation}
Thus we obtain the 1-error correcting
code of parameters exactly as given in Eq.(\ref{d}). Now we shall
demonstrate that its coding subspace is 50\% larger than the
corresponding optimal stabilizer codes so that our codes are genuine
nonadditive codes that are neither equivalent to some stabilizer
codes under local unitary transformations nor subcodes of some
larger 1-error correcting stabilizer codes of the same length.

The quantum Hamming bound for a 1-error correcting stabilizer code,
e.g., $n-k\ge \lceil\log_2(3n+1)\rceil$ for a stabilizer code
$[[n,k,3]]$, being introduced initially for the non-degenerate
codes, is valid for both degenerate and generate codes of distance 3
and 5 \cite{g1} and of a large enough length \cite{ash}.  In the
case of $n=N_m^a$ we have the quantum Hamming bound $n-k\ge 2m+5$.
This is not enough to prove the nonadditiveness of our codes.
However by working out analytically the linear programming bound we
have

{\bf Theorem} If there exists a stabilizer code $[[N_m^a,k,3]]$,
degenerate or non-degenerate, with
$N_m^a=\frac{2^{2m+5}-5}3+a$ and $m\ge 0$ and $a=0,1$ then $N_m^a-k\ge 2m+6$.

{\bf Proof.}  Given a stabilizer code $[[n,k,d]]$ its weight distributions
$\{A_i\}_{i=0}^n$ is defined by
\begin{equation}
A_i=\frac1{2^{2k}}\sum_{|\omega|=i}|\tr(P\mathcal
E_\omega)|^2 \quad (i=0,1,\ldots,n),
\end{equation}
where the summation is over
all errors supported on $i$ qubits and $P$ is the projector onto the coding
subspace. It is obvious that $A_i\ge0$,
$A_0=1$, and $\sum_iA_i=2^s$ so that $\{A_i/2^s\}$ can be regarded
as  a probability distribution with $s=n-k$. For an arbitrary
function $f(x)$ we denote its average
\begin{equation}
\langle f(x)\rangle\equiv \frac1{2^s}\sum_{i=0}^d{f(i)A_i}.
\end{equation}
In the following we shall formulate a subset of the linear
programming bound for 1-error correcting code, which serves our
purpose perfectly. For a complete set of linear programming bound
see Ref.\cite{cal2,rains1}.

{\em Linear Programming bound} (Restricted) If there exists a
stabilizer code $[[n,k,3]]$ then the following conditions hold true
\begin{subequations}
\begin{eqnarray}
A_1=\langle 3n-4x\rangle,\label{lp1}\\
A_2=\frac 12\langle (4x-3n+1)^2-3n-1\rangle,\label{lp2}\\
\sum_{i=0}^{\lfloor \frac n2\rfloor}A_{2i}\ge 2^{s-1}.\label{lp3}
\end{eqnarray}
\end{subequations}

In the case of $a=0$, i.e., $n=N_m^0$ with $m\ge0$ we introduce a nonnegative function
$f(x)=({3n+1}-4x)^2$ and it is easy to check that as long as $n\ge5$
\begin{subequations}
\begin{eqnarray}
f(0)&=&(3n+1)^2>(3n+5)(3n-7)+16,\\
f(1)&=&(3n-3)^2>4(3n+5),\\
f(2)&=&(3n-7)^2>2(3n+5)+16.
\end{eqnarray}
\end{subequations}
If there exists a stabilizer code $[[n,k,3]]$ then
Eqs.(\ref{lp1}-\ref{lp3}) must hold. As a result
\begin{subequations}
\begin{eqnarray}
&\langle f(x)\rangle=3n+1+4A_1+2A_2,&\\
&\displaystyle 16+16A_2+\sum_{i=2}^{\lfloor \frac
n2\rfloor}f(2i)A_{2i}\ge16\sum_{i=0}^{\lfloor \frac
n2\rfloor}A_{2i}\ge 8. 2^{s},&
\end{eqnarray}
\end{subequations}
where we have used that fact that $f(2i)\ge16$ since $\frac{3n+1}4$, the unique
zero of $f(x)$,  is an odd integer. Putting all these pieces
together we obtain
\begin{eqnarray}
&&{2^s}\langle f(x)\rangle =\sum_{i=0}^nf(i)A_i\cr &\ge &
f(0)+f(1)A_1+f(2)A_2+\sum_{i=2}^{\lfloor \frac
n2\rfloor}f(2i)A_{2i}\cr&\ge&
f(0)-16+f(1)A_1+(f(2)-16)A_2+8.2^{s}\cr &>&
(3n+5)(3n-7+4A_1+2A_2)+8.2^{s}\cr &=&(3n+5)\langle
f(x)-8\rangle+8.2^{s},
\end{eqnarray}
in which the strict inequality comes from the $f(0)$ term. Taking into account of $\langle
f(x)\rangle >8$ we obtain $2^s>3n+5$, i.e., $n-k\ge 2m+6$.

In the case of $a=1$, i.e., $n=N_m^1$ with $m\ge 0$ we define $
g(x)=({3n+2}-4x)({3n-2}-4x)$ which is nonnegative on integers
because $\frac{3n+2}4$ is an integer. It is obvious that as long as
$n\ge 5$ we have $g(i)> 2(3n+2)$ for $i=1,2$ and most importantly
$g(0)>(3n+2)(3n-4)$. If there exists a stabilizer code $[[n,k,3]]$
then Eqs.(\ref{lp1}-\ref{lp3}) must hold, which leads to $\langle
g(x)\rangle=3n-4+2A_1+2A_2$. As a result we have
\begin{eqnarray}
{2^s}\langle g(x)\rangle&\ge& g(0)+g(1)A_1+g(2)A_2\cr
&>&(3n+2)(3n-4+2A_1+2A_2)\cr &=&(3n+2)\langle g(x)\rangle,
\end{eqnarray}
in which the strict inequality sign is due to the $g(0)$ term. Since
$\langle g(x)\rangle>0$ we have $2^s> 3n+2$, i.e., $n-k\ge 2m+6$.
 {}\hfill $\square$

It should be noted that the optimal stabilizer codes of parameters
as given in Eq.(\ref{os}) exist and construct is already given  by
the stabilizers in Table I with the stabilizers acting on qubits
$V_0$ or $V_1$ being replaced by 6 stabilizers of the pure optimal
stabilizer codes $[[9,3,3]]$ or $[[10,4,3]]$.

The stabilizer-like structures of our codes simplify significantly
the encoding and decoding procedures. Let us suppose we have already
the  encoding and decoding circuits for the codes $\mathbb
D_{(0,a)}$ and for the Gottesman's codes. With some additional
controlled-not gates in front of the encoding circuits of these
individual codes we obtain the encodings of our codes. To decode we
have only to check at first the first $2m$ generators in Table I,
from which we can be sure wether the errors happen on some U-block
or not. If yes we use the decodings for Gottesmans codes and if not
then we have only to decoding $\mathbb D_{(0,a)}$ with the detailed
circuit in the case of $a=0$ being given in \cite{yu2}.

We acknowledge the financial support of NNSF of China (Grant No.
90303023, 10675107, and 10705025) and the A*STAR grant
R-144-000-189-305.

Note added. On finishing the paper another infinite family of
genuine nonadditive codes has been reported in \cite{grsl2}.


\begin{thebibliography}{99}
\bibitem{shor1} P. W. Shor, Phys. Rev. A,
{\bf 2}, 2493 (1995).
\bibitem{ben}C. H. Bennett, D. P. DiVincenzo, J. A. Smolin, andW. K.Wootters,
Phys. Rev. A 54, 3824 (1996).
\bibitem{ste1} A. Steane, Phys. Rev. Lett. {\bf 77}, 793 (1996).
\bibitem{knill}E. Knill and R. Laflamme,
Phys. Rev. A {\bf 55}, 900 (1997).
\bibitem{g2} D. Gottesman, Phys. Rev. A {\bf 54} 1862 (1996).
\bibitem{g1} D. Gottesman, arXive: quant-ph/9705052.
\bibitem{cal1} A. Calderbank, E.
Rains, P. Shor, and N. Sloane, Phys. Rev. Lett. {\bf 76}, 405
(1997).
\bibitem{cal2} A. Calderbank, E. Rains, P. Shor, and N. Sloane, IEEE Trans. Inform. Theory,
{\bf  44}, 1369 (1998).
\bibitem{rains} E.M. Rains, R. H. Hardin, P.W. Shor, and N.J.A. Sloane,
Phys. Rev. Lett. {\bf 79}, 953 (1997).
\bibitem{rainsa} E.M. Rains, IEEE Trans. Inf. Theory {\bf 45}, 266 (1999).
\bibitem{smolin} J.A. Smolin, G. Smith and S. Wehner, Phys. Rev. Lett. {\bf 99}, 130505 (2007).
\bibitem{yu2}S. Yu, Q. Chen, and C.H. Oh, arXiv: 0709.1780v1 [quant-ph]
\bibitem{yu1}S. Yu, Q. Chen, C.H. Lai, and C.H. Oh, Phys. Rev. Lett.
{\bf 101}, 090501 (2008)
\bibitem{grsl}M. Grassl and M. Roetteler,
Proc. 2008 IEEE Int. Symp. on Inf. Theory (ISIT 2008), 300 (Toronto,
Canada, July 2008).
\bibitem{zb}A. Cross, G. Smith, J. Smolin, and B. Zeng, IEEE Trans.
Inf. Theory {\bf 55}, 433 (2009).
\bibitem{g3} D. Gottesman, arXive: quant-ph/9607027.
\bibitem{rains1} E. Rains, IEEE Trans. Inform.
Theory {\bf  44}, 1388 (1998); ibid, IEEE Trans. Inform. Theory {\bf
45}, 2361 (1999).
\bibitem{shor2}P. Shor and R.
Laflamme, Phys. Rev. Lett. {\bf 78}, 1600 (1997).
\bibitem{ash}
A. Ashikhmin and S. Litsyn, IEEE Trans. Inform. Theory {\bf 45},
1206 (1999).
\bibitem{werner}D. Schlingemann and R.F. Werner, Phys. Rev. A {\bf 65}, 012308
(2001).
\bibitem{graph} M. Hein, J. Eisert, and H.J. Briegel, Phys. Rev. A {\bf 69}.
062311(2004).
\bibitem{grsl2} M. Grassl, P. Shor, G. Smith, J. Smolin, and B. Zeng, arXiv: 0901.1319 [quant-ph].
\end{thebibliography}
\end{document}